\documentstyle[12pt,aaspp4]{article}

\lefthead{Magnetic Energy Spectra in Active Regions}
\righthead{Abramenko}

\begin{document}

\title{Magnetic Energy Spectra in Solar Active Regions}

\author{Valentyna Abramenko and Vasyl Yurchyshyn}
\affil{Big Bear Solar Observatory, 40386 N. Shore Lane, Big Bear City, CA
92314}

\begin{abstract}

Line-of-sight magnetograms for 217 active regions (ARs) of different flare rate
observed at the solar disk center from January 1997 until December 2006 are
utilized to study the turbulence regime and its relationship to the flare
productivity. Data from {\it SOHO}/MDI instrument recorded in the high
resolution mode and data from the BBSO magnetograph were used. The turbulence
regime was probed via magnetic energy spectra and magnetic dissipation spectra.
We found steeper energy spectra for ARs of higher flare productivity. We also
report that both the power index, $\alpha$, of the energy spectrum, $E(k) \sim
k^{-\alpha}$, and the total spectral energy $W=\int E(k)dk$ are comparably
correlated with the flare index, $A$, of an active region. The correlations are
found to be stronger than that found between the flare index and total unsigned
flux. The flare index for an AR can be estimated based on measurements of
$\alpha$ and $W$ as $A=10^b (\alpha W)^c$, with $b=-7.92 \pm 0.58$ and $c=1.85
\pm 0.13$. We found that the regime of the fully-developed turbulence occurs in
decaying ARs and in emerging ARs (at the very early stage of emergence).
Well-developed ARs display under-developed turbulence with strong magnetic
dissipation at all scales.

\end{abstract}

\keywords{Sun: activity - Sun: flares - Sun: photosphere - Sun: surface
magnetism - Physical Data and Processes: magnetic fields - 
 Physical Data and Processes: turbulence }

\section { Introduction}

Existing methods to predict flare activity of ARs are not numerous (e.g.,
Abramenko et al. 2002; Falconer et al. 2002, 2003, 2006; Leka \& Barnes 2003,
2007; McAteer et al. 2005; Schrijver 2007; Georgoulis \& Rust 2007; 
Barnes \& Leka 2008; see review
by McAteer et al. 2009) and majority of them are based on the first-order
statistical moments of the magnetic field (test parameters are derived from
$B$). At the same time, Leka \& Barnes (2007) noted that exploration of higher
order statistical moments seems beneficial given the non-linear nature of a
flaring process. 

Earlier we presented results of a small statistical study (Abramenko 2005)
focused on the relationship between flare productivity of solar active regions
(ARs) and the power-law index, $\alpha$, of magnetic energy spectrum,  $E(k)
\sim k^{-\alpha}$. This magnetic energy spectrum technique is based on the
second-order statistical moment of a 2D field, and it shows distribution of
energy over spatial scales. The study was based on 16 ARs observed
predominantly during 2000 - 2003 with the {\it SOHO}/MDI instrument (Scherrer et
al. 1995) performing in the high resolution (HR) mode. The findings were
promising: flaring ARs were reported to display steeper power spectra with the
power index exceeding the magnitude of 2. Flare-quiet ARs exhibited a
Kolmogorov-type spectrum with $\alpha$ close to 5/3 (Kolmogorov 1991, referred
hereinafter K41).

Power spectra calculations, based on the technique described in Abramenko (2005)
will be a part of the pipeline system designed to process real-time data flowing
from Helioseismic and Magnetic Imager\footnote{http://hmi.stanford.edu/} that
operates on board of the {\it Solar Dynamics
Observatory}\footnote{http://sdo.gsfc.nasa.gov/}. Here we evaluate performance
of the method based on a large data set. We report that both the magnetic energy
stored in large-scale structures  and the magnetic energy cascaded by turbulence
at small-scale structures (below 10 Mm), are comparably correlated with flare
productivity (Section 3). From the magnetic energy spectra we also determined at
what spatial scales the bulk of the magnetic energy dissipation occurs (Section
4). This allowed us to make an inference about the characteristics of the
turbulent regime in ARs, which may be useful as a constraint criterion
for the MHD modeling of ARs.

\section { Data }

We selected 217 active regions measured near at the solar disk center (no
further than 20 degrees away from the central meridian), so that the projection
effect can be neglected. The set covers the period between January 1997 and
December 2006. All the ARs displayed nonzero flare activity, i.e., at least
one GOES flare flare was produced by an AR during its passage across the solar
disk. Given typical average flux densities in active regions, smooth power
spectra are not usually obtained for active regions with unsigned magnetic flux
less than 10$^{22}$ Mx. Therefore, we required that each AR should possess
unsigned magnetic flux that exceeds this threshold value.

For majority of ARs (215), MDI/HR magnetograms (pixel size of 0.6$^{''}$) were
utilized. For two extremely flare-productive ARs only Big Bear Solar Observatory
(BBSO)
data, obtained at very good seeing conditions with the Digital Magnetograph
(DMG, Spirock 2005, pixel size of 0.6$^{''}$) were available. As we have shown
earlier (Abramenko et al. 2001) magnetic energy spectra calculated for the same
AR from MDI/HR and BBSO/DMG magnetograms agree very well at scales larger than 3
Mm. This allows us to use different data without risk of skewing the
correlation. From MDI full disk magnetograms we determined the trend in the
total unsigned flux in each AR during a 3-day time interval centered at the time
of MDI/HR magnetogram acquisition. When AR flux variations did not exceed
$\pm$10\% of the mean value we classified the AR as a stable, well-developed AR.
ARs displaying larger monotonous changes of the flux were classified as emerging
or decaying. Unipolar sunspots were detected by visual assessment.

Flare productivity of an AR was measured by the flare index, $A$, introduced
in Abramenko (2005). Since the X-ray classification of solar flares (X, M, C,
and B) is based on denary logarithmic scale, we can define the flare index as
\begin{equation}
A=(100 S^{(X)}+10 S^{(M)}+ S^{(C)}+0.1 S^{(B)})/t.
\label{A}
\end{equation}
Here, $S^{(j)}$ is the sum of all GOES flare magnitudes in the $j$-th X-ray
class:
\begin{equation}
S^{(j)}=\sum_{i=1}^{N_j} I_i^{(j)},
\label{AA}
\end{equation}
where $N_j=N_X, N_M, N_C$ and $N_B$ are the numbers of flares of X, M, C and B
classes, respectively, that occurred in a given active region during its passage
across the solar disk that is represented by the time interval $t$ measured in
days. $I_i^{(j)}=I_i^{(X)}, I_i^{(M)}, I_i^{(C)}$ and $I_i^{(B)}$ are GOES
magnitudes of X, M, C ans B flares. The interval $t$ was taken to be 27/2 days
for the majority of the ARs with exception of emerging ones. In general, those
ARs that produced only C-class flares have flare indices smaller than 2, whereas
several X-class flares will result in the flare index exceeding 100. The highest
flare index of 584 was registered for NOAA AR 10486 which, unfortunately, was
not included in the analysis because it was outside of the MDI/HR field of view.

\section { Magnetic Energy Spectra  }

Photospheric plasma is thought to be in a turbulent state, subject to the
continuous sub-photospheric convection. It can be analyzed with the energy
spectrum (which is frequently referred as a power spectrum). The energy spectrum
is based on the second-order statistical moments of a field and it is very
useful to probe the inter-scale energy cascade and dissipation regimes (see,
e.g., Monin \& Yaglom 1975; Biskamp 1993).

\begin{figure}[!h] \centerline {
\epsfxsize=3.0truein
\epsffile{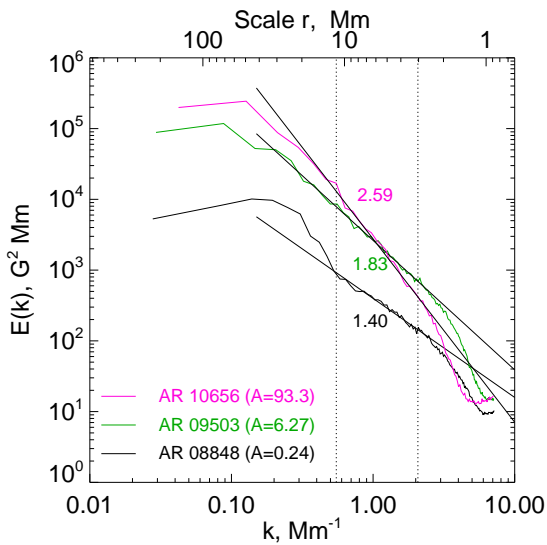}}
\caption{\sf Magnetic energy spectra, $E(k)$, shown for three ARs with different
flare index, $A$. Vertical dotted segments mark the linear interval between
3-10~Mm, where the the best linear fits (thin lines) were calculated. The power
index, $\alpha$, defined as $E(k) \sim k^{-\alpha}$ and measured as a slope of
the best fit for each spectrum is shown.}
\label{fig1}
\end{figure}

Our method to calculate the 1D angle-integrated energy spectrum, $E(k)$, of a 2D
structure (e.g., a line-of-sight magnetogram, $B_z(x,y)$) was described in
Abramenko et al. (2001) and later improved in Abramenko (2005). The most
reliable scale interval to derive the energy cascade in the spectrum is 3 -
10~Mm when utilizing MDI/HR and BBSO data (Abramenko et al., 2001). At scales
smaller than 2-3~Mm, the influence of insufficient resolution is substantial,
while the contribution from large sunspots at scales exceeding 10~Mm
contaminates the slope of the turbulent cascade in the spectrum. For all the
ARs, we used the above linear interval to measure the power index, $\alpha$, of
the magnetic energy spectrum, $E(k) \sim k^{-\alpha}$, where $k$ is a wave
number inversely proportional to the spatial scale, $r= 2\pi/k$. The index was
computed from the best linear fit in the double logarithmic plot via the
IDL/LINFIT routine by minimizing the $\chi$-square error statistic.
Figure \ref{fig1} shows energy spectra for three ARs of different flare
productivity which is evident from different steepness of the slopes. The
well-defined power-law energy cascade interval is observed at 3-10~Mm in all
cases. 
\begin{figure}[!h] \centerline { \epsfxsize=6.0truein
\epsffile{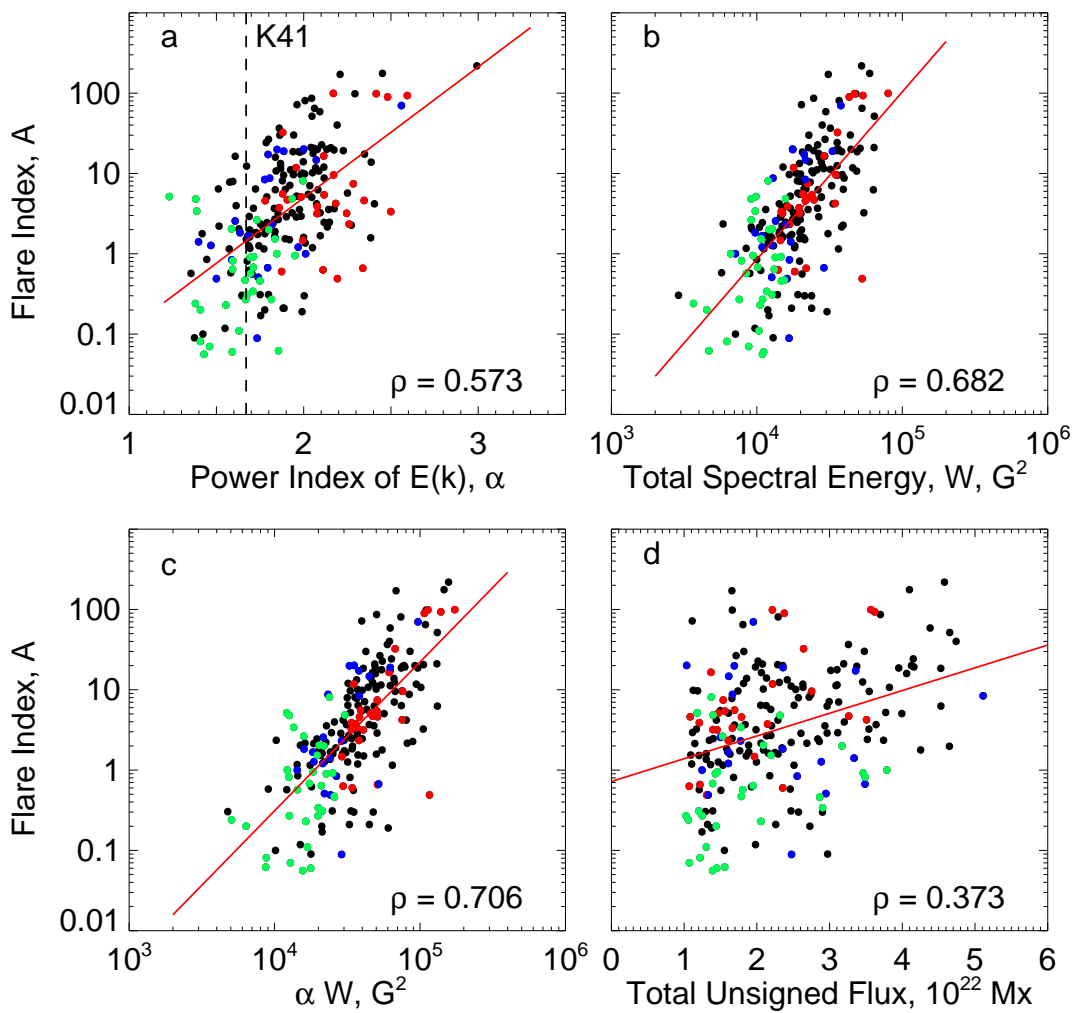}} \caption{\sf  The flare index, $A$ plotted
versus: {\it a} -  
the power index, $\alpha$, of the magnetic energy spectrum (the vertical dashed
segment marks the power index of the fully-developed Kolmogorov-type turbulence
of $\alpha=5/3$); {\it b} - the total
spectral energy, $W$; 
{\it c} - the power index
weighted by the total spectral energy, $W$;  
{\it d} - the total unsigned flux. 
The solid lines show the best linear
fits to the data points. The correlation coefficients, $\rho$, are shown. Black,
red, blue and green circles mark well-developed, emerging, decaying ARs and
unipolar sunspots, respectively.}
\label{fig2}
\end{figure}

The relationship between the power and flare indexes is shown in Figure
\ref{fig2}{\it a}. Positive correlation between these parameters ($\rho=0.57$
with the 95\% confidence interval of 0.478 - 0.66 according to the Fisher's
Z-transformation statistical test of
significance\footnote{http://icp.giss.nasa.gov/education/statistics/page3.html})
allows us to conclude that, in general, steeper spectra are associated with
higher flaring rate. Unipolar active regions (green circles) form a separate
low-flaring/shallow-spectrum subset. Emerging, decaying and stable ARs are
distributed more or less uniformly over the diagram.

Comparison of the spectra in Figure \ref{fig1} also suggests that not only the
slope, but the amplitudes of a spectrum may be related to flare productivity.
Thus, we calculated two additional parameters from the spectra. The first
parameter, $\langle E(k_{low}) \rangle$, characterizes the amplitude of a
spectrum at low wave numbers and is derived as $E(k)$ averaged over the interval
ranging from the smallest $k$ to $k=2\pi/ 10 $ Mm$^{-1}$. The correlation
coefficient between this parameter and the flare index $A$ is 0.65 with the 95\%
confidence interval of 0.57 - 0.72. The second parameter is the total spectral
energy, $W = \int E(k) dk$, where the integration is performed over all
wavenumbers where the $E(k)$ is non-zero. On average, 85$\pm 5$ \% of the total
spectral energy is concentrated at scales larger than 10 Mm, so that $W$ may be
considered as a measure of energy accumulated in large-scale structures of an
AR. The correlation between $W$ and the flare index $A$ (see Figure
\ref{fig2} {\it b}) was found to be somewhat
higher: $\rho=0.68$ with the 95\% confidence interval of 0.60 - 0.75.

Note that, $W$ can also be computed directly from $B_z^2$ according to
Parseval's theorem (e.g., Kammler 2000). However, computing $W$ via integration
of a Fourier transform has an advantage because the integration over an annulus
also acts as a noise filter by leaving out high-frequency corners in the 2D wave
number box. When analyzed data come from MDI/HR, the difference in derived
values is less than 0.5\%. However, the difference increases when ground-based
data are analyzed. When a mixed data set is used (space and ground bases
observations), the values of $W$ derived via integration of the Fourier
transform are more consistent. Therefore, the total spectral energy used here
was calculated using the integration approach.

So, the power index of the spectrum, the total spectral energy and the averaged
large-scale amplitude are comparably correlated with the flare index. We may
conclude that both the turbulent energy cascade and the presence of large-scale
structures in an AR are relevant to flare activity.

For the purpose of prediction of AR's flare productivity, we weighted the total
spectral energy, $W$, and the power index, $\alpha$, to capture the contribution
of both small-scale and large-scale effects (Figure \ref{fig2}{\it c}). The
correlation coefficient in this case increased to $0.71$ with the 95\%
confidence interval of 0.63 - 0.77. The flare index can the be then fitted
assuming
\begin{equation}
A = 10^b (\alpha W)^c,
\label{A}
\end{equation}
where $b=-7.92 \pm 0.58$ and $c=1.85 \pm 0.13$, with the reduced $\chi^2=0.29$.

In general, the most powerful flares occur in strong, large ARs with
considerable amount of the magnetic flux (e.g., Barnes \& Leka 2008). And yet,
the total unsigned flux is only very weakly correlated to the flare index
(Figure \ref{fig2}{\it d}, the correlation coefficient is 0.37 with the 95\%
confidence interval of 0.25 - 0.49). This is understandable when we recall
previous studies showing that not only the magnetic flux but rather {\it
complexity} of the magnetic structure is relevant to the flaring rate (Sammis et
al. 2000; Abramenko et al. 2002; Falconer et al. 2002, 2003, 2006; Leka \&
Barnes 2003, 2007; McAteer et al. 2005; Schrijver 2007; Georgoulis \& Rust
2007).

\section{\bf Magnetic Dissipation Spectra }

As it follows from Figure \ref{fig2}{\it a}, the majority of ARs
(especially those of high flare activity) display energy spectra steeper than
the Kolmogorov-type spectrum. To infer a physical meaning of this result,
we will analyze the magnetic energy dissipation rates and magnetic dissipation
spectra.

Magnetic energy dissipation rate is related to the presence of electric
currents, i.e., $ \langle \varepsilon \rangle \sim \eta \langle {\bf
j}^2\rangle$ (e.g., Biskamp 1993). In MHD models of turbulence, dissipative
structures are visualized via (squared) currents (e.g., Biskamp \& Welter 1989;
Biskamp 1996;  Schaffenberger et al. 2006; Pietarila Graham et al. 2009;
Servidio et al. 2009) which appear in 2D images to be predominantly located
along magnetic field discontinuities frequently referred to as current sheets.
From 2D MHD modeling, Biskamp and Welter (1989) found that when the magnetic
Reynolds number (which is the ratio of characteristic values of advection terms
to the magnetic diffusivity and quantifies the strength of advection relative to
magnetic diffusion) is low, these current sheets are extended and rare, and they
become shorter and more numerous as Reynolds number increases. Thus, the
magnetic dissipation spectrum represents the distribution of dissipative
structures over many spatial scales, and it is a reasonable proxy for statistics
of current structures in an AR.

\begin{figure}[!h] \centerline { \epsfxsize=6.0truein
\epsffile{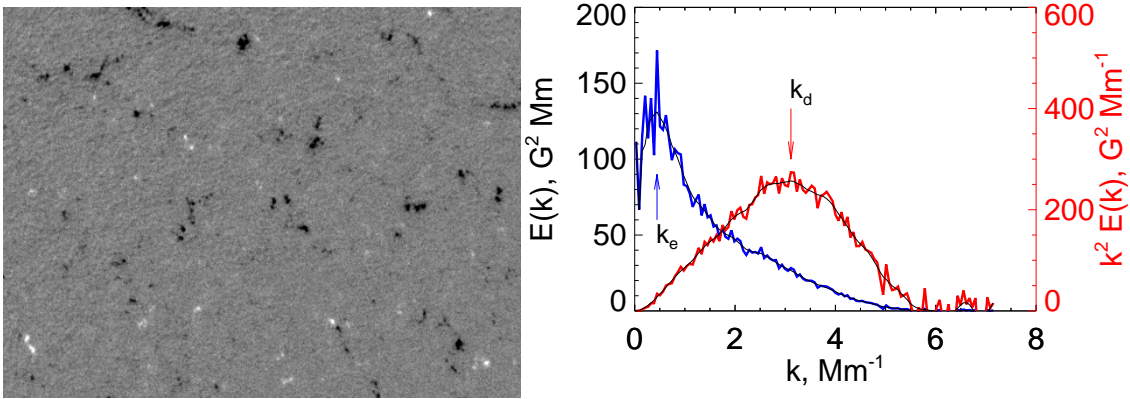}} \caption{\sf MDI/HR magnetogram for a quiet sun area
recorded on 2001 June 5/13:07 UT ({\it left}) and corresponding spectra ({\it
right}): the energy spectrum $E(k)$ ({\it blue curve and right axis}) and
dissipation spectrum, $k^2E(k)$ ({\it red curve, left axis}). The image size is
$ 266 \times 202$ arcsec. The magnetogram is scaled from -150 G to 150G. The
arrows $k_e$  and $k_d$ mark the maxima of the energy ($k_e$) and dissipation
($k_d$) spectra.  Positions of the maxima were derived by 5 point box car
averaging ({\it black curves}). The maxima of the energy and dissipation spectra
are distinctly separated in the wavenumber space. }
\label{figQS}
\end{figure}

The magnetic dissipation spectrum allows us probe the state of
turbulence. For fully developed turbulence (K41, high Reynolds number), the bulk
of magnetic energy dissipation occurs at small scales, $k_d$, whereas the the
energy input occurs at large scales, $k_e$, (Figure \ref{figQS}) and energy
cascades from large to small scales without losses. When the energy input
interval and the dissipation interval overlap, dissipation occurs at
intermediate scales, along the cascade. This condition occurs in the state of
under-developed turbulence (low Reynolds number), when large-scale structures
might interfere with the turbulent cascade at small scales. It is a challenge to
model such a field because no K41 simplifications are applicable.

The magnetic energy dissipation spectra are defined as (Monin \& Yaglom 1975,
Biskamp 1993):
\begin{equation}
E_{dis}(k) = 2\eta k^2E(k),
\label{Edis}
\end{equation}
where $\eta$ is the magnetic diffusivity coefficient. (Note that $E$ and
$E_{dis}$ in Eq. \ref{Edis} have different dimensions.) Then the rate of
magnetic energy dissipation normalized by the magnetic diffusivity can be
derived as (Biskamp 1993):
\begin{equation}
\langle \varepsilon \rangle/\eta = 2{\int_0^{\infty}}  k^2 E(k) dk.
\label{epsilon}
\end{equation}

From observations we can derive function $k^2E(k)$, which is proportional to the
dissipation spectrum under an assumption that  $\eta$ is uniform over the AR
area. In our case both $k^2E(k)$ and  $\langle \varepsilon \rangle/\eta$ are
associated with dissipation of the $B_z$ component only.

\begin{figure}[!h] \centerline {\epsfxsize=5.5truein
\epsffile{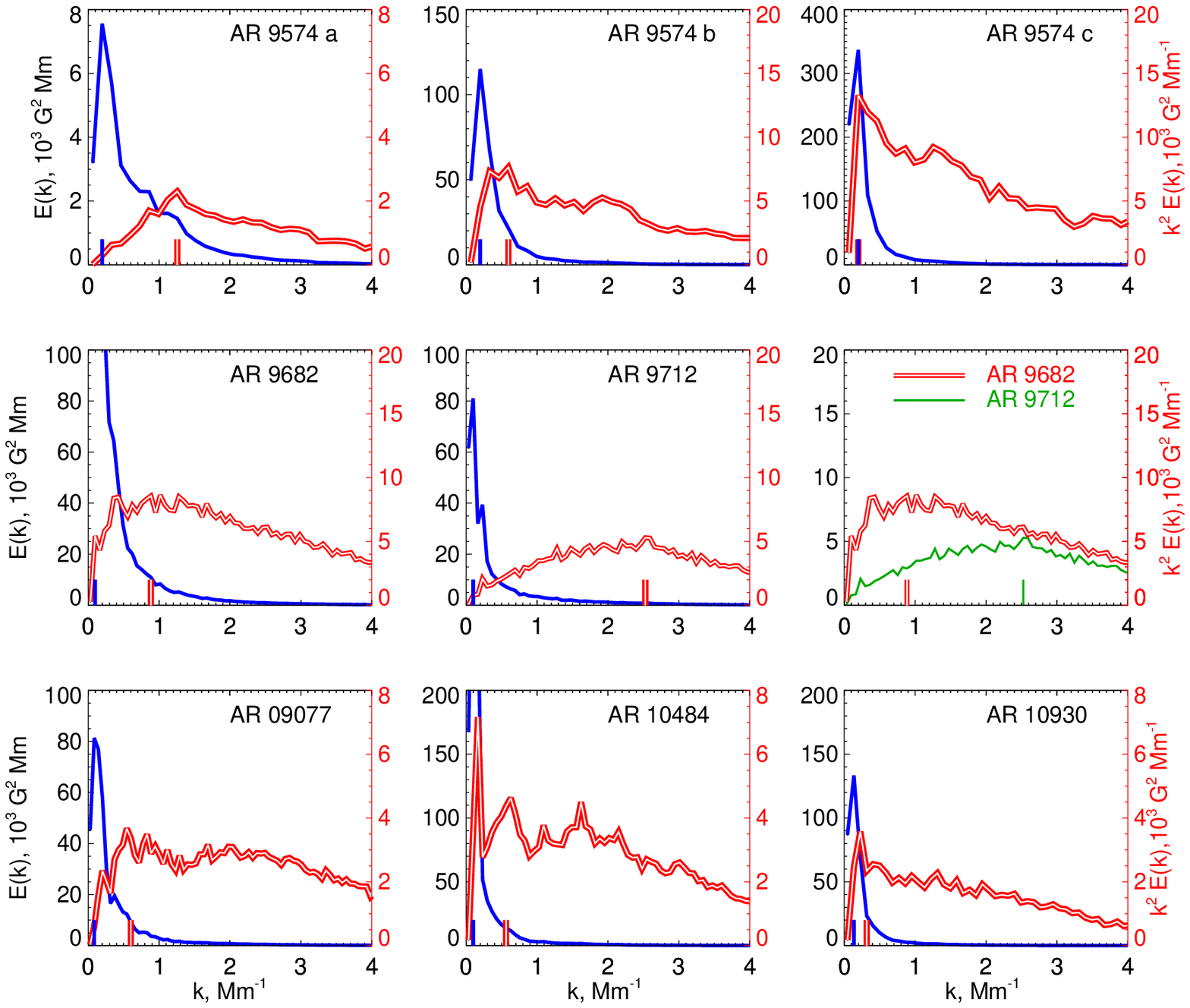}}
\caption{\sf {\it Top - } energy spectra, $E(k)$ ({\it blue lines}), and
dissipation spectra, $k^2E(k)$ ({\it double red lines}), plotted for an emerging
AR. Panels {\it a, b} and {\it c} correspond to three consecutive moments during
the emergence. Vertical blue (red) bars mark the maximum of the energy
(dissipation) spectrum. Blue bars correspond to $k_e$ and red bars correspond to
$k_d$. As the active region emerges, $k_d$ shifts toward the smaller
wavenumbers. {\it Middle} - energy and dissipation spectra for a decaying
magnetic complex NOAA ARs 9682 and 9712. The right panel shows a superposition
of the dissipation spectra at the well-developed ({\it double red line}) and
decaying ({\it solid green line}) state of the magnetic complex. As the magnetic
fields decay, $k_d$ shifts toward small scales (larger wavenumbers). {\it Bottom
- } energy spectra and dissipation spectra for three well-developed ARs. The
spectra are overlapped for each case.}
\label{fig4}
\end{figure}

We calculated $k^2E(k)$ spectra for all ARs in our data set. Typical examples
are shown in Figure \ref{fig4}. At the early stage of development of emerging
ARs, the separation distance ($k_d - k_e$) is largest, which is similar to the
fully-developed turbulence conditions seen in quiet Sun (see Figure
\ref{figQS}). Later on this distance decreases as $k_d$ shifts toward smaller
wavenumbers (larger scales), so that the intervals of energy and dissipation
become exceedingly overlapping. This implies formation of large-scale
dissipative structures. Decaying magnetic complexes show quite opposite behavior
(Figure \ref{fig4}, middle row). Well-developed ARs (bottom row in Figure
\ref{fig4}) show a significant overlap of the energy and dissipation intervals
suggesting that, to the contrary of the fully developed turbulence
phenomenology, significant dissipation takes place at all spatial scales.
Thus, for the majority of well-developed ARs, one should expect a state
of under-developed turbulence in the photosphere with the dissipation of the
magnetic energy at all observable spatial scales.

We then compared the magnitudes of $\langle \varepsilon \rangle/\eta$ to the
flare index, $A$. Their correlation turned out to be positive with
$\rho=0.53$ with the 95\% confidence interval of 0.43 - 0.62. This indicates
that the rate on magnetic energy dissipation in the photosphere is relevant to
flare activity.

\section {Conclusions and Discussion}

In this study we analyzed second-order statistical moments of solar magnetic
fields of 217 active regions observed with the MDI instrument in high
resolution mode during the 23rd solar cycle.

The angle-integrated magnetic energy spectra of solar ARs display a well-defined
power-law region, which indicates the presence of a turbulent non-linear energy
cascade. The power index, $\alpha$ measured at 3-10~Mm scale range was found to
be well correlated with the flare index, $A$ (correlation coefficient,
$\rho=0.57$). This results further supports our previous findings based on
only 16 ARs (Abramenko 2005). The power indices range between 1.3 and 3.0, with
the majority of ARs having the power index in the range of $1.6 - 2.3$. No
particular preference for the classical 5/3 index was found. These values are in
a surprising agreement with recent numerical simulations of decaying MHD
turbulence (Lee et al. 2009). The model results showed that in equivalent
initial magnetic configurations different types of spectra (from $k^{-3/2}$ to
$k^{-2}$) may emerge depending on the intrinsic non-linear  dynamics of the
flow.

The total spectral energy, $W=\int E(k) dk$, is found to be well correlated with
the flare index ($\rho=0.68$), while spectral energy weighted by the power
index shows the strongest correlation  to the flare index ($\rho=0.71$), which
allowed us to determine an empirical description of this relationship:
$A=10^b(\alpha W)^c$, where $b=-7.92 \pm 0.58$ and $C=1.85 \pm 0.13$.

Combined analysis of magnetic energy and magnetic dissipation spectra showed
that in majority of well-developed ARs, the turbulent energy cascade is
augmented by magnetic energy dissipation at all scales. We thus argue that a
state of under-developed turbulence exists in the photosphere of mature ARs.

The magnetic energy dissipation rate, $\langle \varepsilon \rangle/\eta$,
correlates with the flare productivity in the same degree as the power index
does ($\rho=0.53$). As long as energy dissipation rate is proportional to
electric currents squared, we argue that the presence of currents is relevant to
flare productivity. Also, good correlation between the energy dissipation rate
and the flaring rate is in agreement with earlier reports by Schrijver et al.
(2005).

It is known from direct calculations based on vector-magnetograms that electric
currents are ubiquitous in ARs (see, e.g., Abramenko et al. 1991; Leka et al.
1993, 1996; Pevtsov et al. 1994; Abramenko et al. 1996; Wheatland 2000; Zhang
2002; Schrijver et al. 2005; Leka \& Barnes 2007; Schrijver et al. 2008;
Schrijver 2009). We found here that photospheric magnetic fields are in a state
of under-developed turbulence when both the energy cascade and energy
dissipation at all scales are present in the system. We therefore, arrive at a
conclusion that both large and small scale dissipative structures (currents) are
relevant to flaring.

On the other hand, Fisher et al. (1998) found no correlation between
photospheric currents and soft X-ray luminosity of ARs. This was also noted, but
not discussed, by Schrijver et al. (2005). We suggest that this apparent
controversy is due to different nature of flares and the soft X-ray emission. We
may consider flares as sporadic explosive events caused by strongly non-linear
dynamics relevant to large- and small-scale magnetic discontinuities (Falconer
et al. 2002, 2003, 2006, see also Schrijver 2009), whereas soft X-ray emission
reflects a more stationary and homogeneous process of coronal heating rather
related to ubiquitous small scale discontinuities formed {\it in-situ} (Klimchuk
2006).

We would also like to note that the power law of the magnetic energy spectra,
explored in this paper, shell not be confused with the power law found in the
distribution of the magnetic flux in flux concentrations reported recently by
Parnell et al. (2009). At first sight, both of them characterize the {\it
structure} of the magnetic field. However, they address different physical
consequences of the magnetic field structuring. The power law of the magnetic
energy spectrum represents distribution of magnetic energy, $B^2$, over spatial
scales and quantifies turbulence in an AR. Here, the smallest magnetic elements
are represented by the tail of the spectrum usually associated with low spectrum
amplitudes. The power law found in distribution of magnetic flux represents
frequency (abundance) of magnetic elements of different sizes and implies a
unique mechanism of formation of magnetic flux concentrations (say,
fragmentation process, see Abramenko \& Longcope (2005) for more discussion).
The smallest magnetic elements are the most frequent and they are represented by
the highest amplitudes of the distribution.

Modern computational capabilities allow us to develop MHD models which take
into account turbulent regime and turbulent dissipation (e.g., Lionello et al.
2010; Klimchuk et al. 2010). Therefore, diagnostics of turbulence derived from a
large uniform data set is essential for constructing and restraining of these
models.

We are grateful to anonymous referee for criticism and useful comments allowing
to improve substantially the manuscript. SOHO is a project of international
cooperation between ESA and NASA. This work was supported by NSF grant
ATM-0716512 and NASA grants NNX08AJ20G and NNX08AQ89G.

{}

\end{document}